\documentclass[12pt]{article}
\usepackage{latexsym,amsmath,amssymb,epsfig,graphicx}
\newcommand{\be}{\begin{equation}}
\newcommand{\ee}{\end{equation}}
\newcommand{\bea}{\begin{eqnarray}}
\newcommand{\eea}{\end{eqnarray}}
\newcommand{\kma}{\; ,}
\newcommand{\pkt}{\; .}
\newcommand{\diag}{{\rm diag}}
\newcommand{\calm}{{\cal M}}
 
\newcommand{\calg}{{\cal G}}
\newcommand{\call}{{\cal L}}

\newcommand{\bfx}{{\bf x}}
\newcommand{\bfX}{{\bf X}}
\newcommand{\bfH}{{\bf H}}
\newcommand{\bfP}{{\bf P}}
\newcommand{\bfY}{{\bf Y}}

\newcommand{\tr}{{\rm Tr}}
\newcommand{\eqn}[1]{(\ref{#1})}

\begin{document}
\begin{titlepage}
\begin{flushright}
hep-th/yymmnn \\
December 2010
\end{flushright}
\vspace{8mm}
\begin{center}
{\Large \bf
One-loop corrections to the Nielsen-Olesen vortex: collective oscillations.}
\\\vspace{8mm}
{\large  J\"urgen Baacke\footnote{e-mail:~
juergen.baacke@tu-dortmund.de}} \\
{  Fakult\"at Physik, Technische Universit\"at Dortmund \\
D - 44221 Dortmund, Germany
}\\
\vspace{4mm}

\bf{Abstract}
\end{center}

We connect the translation modes of the instanton in the
two-dimensional Abelian Higgs model with local translations
of the vortex of the related  model in (3+1) dimensions,
the Nielsen-Olesen vortex. In this context these modes describe
collective oscillations of the string. We construct the wave
function of this mode and we derive, via a virial theorem, an
effective action for these oscillations, which is consistent 
with the action constructed by Nielsen and Olesen using general
arguments. We discuss some aspects of renormalization, based
on a recent computation of one loop corrections to 
string tension of the vortex.
\end{titlepage}

%*******************************************************Introduction

\section{Introduction}

A long time ago Nielsen and Olesen \cite{Nielsen:1973cs}
discussed the vortex solution
of the (3+1) dimensional Abelian Higgs model as a possible model
for strings. In the same publication the authors discuss the
effective action for the transverse collective oscillations of the 
vortex which they find to be equivalent to the
Nambu-Goto action of string theory (see e.g. \cite{Green:1987sp}).
Their derivation is based on a general consideration of Lorentz
transformation properties of such oscillations.
We here derive the nonrelativistic limit of 
this effective action from an analysis of
the fluctuations in the underlying quantum field theory.

For quantum kinks
(see, e.g., \cite{Rajaraman:1982is,Coleman85})
 the collective motion is related to the
 translation mode. It is generated by infinitesimal
displacements of the classical kink solution
 and is a zero mode of the 
fluctuation operator. Its quantization requires a special
approach, by which the zero mode is found to carry
 the kinetic energy of the collective
motion of the kink.

In the Abelian Higgs model in two dimensions  one finds an instanton
solution which describes topological transitions
\cite{Kripfganz:1989vm}. The
fluctuations around the classical solution display two
zero modes which again are related to translation invariance.
The one-loop prefactor for the semiclassical transition rate
is related to the functional determinant of the fluctuation operator.
The zero modes would cause this determinant to be infinite and have
to be eliminated. If handled properly, this elimination 
produces the correct dimension for the transition rate
\cite{Baacke:1994bk,Baacke:2008zx}.

The instanton of the Abelian Higgs model reappears in the
$(3+1)$ dimensional version of the model as the
vortex solution which we will consider here. The vortex solution is identical
to the instanton solution in the transversal $x$ and $y$ coordinates,
 and it is independent on $z$ and of time. The translation of the
classical solution now becomes local, dependent on $z$.
Instead of a collective motion of a quantum kink we have
collective oscillations of the vortex.
The pole at $p_\perp=0$ in the Euclidian Green's function of the 
two-dimensional model becomes a cut in the Green' s function of the
model in four dimensions. In the computation  of one-loop
corrections to the string tension \cite{Baacke:2008sq}
these modes can be included in the same
way as all other fluctuations, in contrast to the
two-dimensional case. Indeed, for the renormalization of
these corrections it is necessary to include the
zero mode contribution. 
Still the translation modes play a special r\^ole, 
and in the present work we will discuss these particular aspects.

The text is organized as follows:
In Sec. \ref{sec:basics} we present the model, the classical vortex solution
and the classical string tension.
In Sec. \ref{sec:flucsandtension} 
we define the fluctuation operator and relate it to the one-loop 
correction to the string tension. Based on the derivation of the
translation mode wave functions in Appendix \ref{app:translationmode}
and using a virial theorem proven in Appendix \ref{app:virialtheorem}
we discuss in Sec. \ref{sec:collectivestringoscillations}
the collective string oscillations and their effective action.
We discuss in Sec. \ref{sec:fluctuationenergy} the r\^ole of the 
zero modes in the computation of the renormalized the string tension. 
Conclusions are presented in Sec. \ref{sec:conclusions}.

%%%%%%%%%%%%%%%%%%%%%%%%%%%%%%%%%%%%%%%%%%%%%%%%%%%%%%%%%%%%%%%%%%%%%%%%%

\section{Basic relations}
\setcounter{equation}{0}
\par
\label{sec:basics}
The Abelian Higgs model in (3+1) dimensions is
defined by the Lagrange density 
\begin{equation}
  {\cal L}=-\frac{1}{4}F_{\mu\nu}F^{\mu\nu}
+\frac{1}{2}(D_\mu\phi)^*D^\mu\phi-\frac{\lambda}
{4}\left(|\phi|^2-v^2\right)^2\; .  
\end{equation}
Here $\phi$ is a complex scalar field and
\bea
F_{\mu\nu}&=&\partial_\mu A_\nu-\partial_\nu A_\mu\kma \\
D_\mu&=&\partial_\mu-igA_\mu 
\pkt\eea 
The particle spectrum consists of Higgs bosons of mass
$m_H^2=2\lambda v^2$ and vector bosons of mass $m_W^2=g^2v^2$.
The model allows for vortex type solutions, representing
strings with a magnetic flux, the Nielsen-Olesen vortices 
\cite{Abrikosov:1957,Nielsen:1973cs,deVega:1976mi}.
The cylindrically symmetric
ansatz for this solution is given, in the singular gauge, by 
\footnote{ We use Euclidean notation for the transverse components, so
$A^\perp_1\equiv A^1=-A_1$ etc. } 
\begin{eqnarray}
A^{{\rm cl}\perp}_i (x,y,z)&=&\frac{\varepsilon_{ij}x^\perp_j} 
{gr^2}\left[A(r)+1\right] \;\;\; i=1,2 \\
\phi^{cl}(x,y,z)&=&vf(r) \; .
\end{eqnarray}
where $r=\sqrt{x^2+y^2}$ and $\varphi$ is the polar angle. 
Furthermore $A_3^{\rm cl}=A_0^{\rm cl}=0$. 
With this ansatz the energy per unit length, or string
tension $\sigma$ takes the form
\begin{eqnarray}
\sigma_{\rm cl}&=&\pi v^2
 \int^{\infty}_{0}\!\!\! dr \left\{\frac{1}{rm_W^2}\left[
\frac{dA(r)}{dr}\right]^2\!\!+r\left[\frac{df(r)}{dr}\right]^2\!\!
+\frac{f^2(r)}{r}\left[A(r)+1\right]^2\!\! \right. \nonumber \\ 
&+&\left. \frac{rm^2_H}{4}\left[
f^2(r)-1\right]^2\!\right\}\pkt
\label{eq:classtens} \end{eqnarray}
The classical equations of motion are given by
\begin{eqnarray}
\left\{\frac{\partial^2}{\partial r^2}+
\frac{1}{r}\frac{\partial}{\partial r}-\frac{\left[A(r)+1\right]^2}{r^2}-
\frac{m^2_H}{2}\left[f^2(r)-1\right]
\right\}f(r)&=&0 \kma \\
\left\{\frac{\partial^2}
{\partial r^2}-\frac{1}{r}
\frac{\partial}{\partial r}-m^2_W f^2(r)\right\}
\left[A(r)+1\right]&=&0
\kma\end{eqnarray}
which are to be solved numerically with
\begin{equation}  \label{rb}
\begin{array}{rcccccr}
A(r)&\stackrel{\scriptscriptstyle{r\to 0}}
{\longrightarrow}& {\rm const.}\cdot r^2                       
&,&A(r)&\stackrel{\scriptscriptstyle{r\to\infty}}
{\longrightarrow}&-1 \kma \\ 
f(r)&\stackrel{\scriptscriptstyle{r\to 0}}
{\longrightarrow}&{\rm const.}\cdot r 
&,&f(r)&\stackrel{\scriptscriptstyle{r\to\infty}}
{\longrightarrow}&1\kma 
\end{array}
\pkt\end{equation}

%%%%%%%%%%%%%%%%%%%%%%%%%%%%%%%%%%%%%%%%%%%%%%%%%%%%%%%%%%%%%%%%%%%%%%%%%

\section{Fluctuation operator and one-loop  string tension}
\label{sec:flucsandtension}
Expanding the gauge and Higgs fields as
\bea
\phi &=& \phi_i^{\rm cl} + \varphi_1 + i \varphi_2
\\
A_\mu&= &A_\mu^{\rm cl} + a_\mu
\eea
the dynamics of the fluctuations is described by the second
order gauge fixed Lagrangian \cite{Baacke:1994bk}
\begin{eqnarray} \label{lag2}
\call^{II}&=&
-a_\mu\frac{1}{2}\left(-\Box-g^2\phi^2\right)
a^\mu\nonumber \\
&&+\varphi_1\frac{1}{2}\left[-\Box+g^2A_\mu A^\mu-
\lambda\left(3\phi^2-
v^2\right)\right]\varphi_1  \nonumber \\  
&&+\,\varphi_2\frac{1}{2}\left[-\Box+g^2A_\mu A^\mu-g^2\phi^2-
\lambda\left(\phi^2-v^2\right)\right]\varphi_2 \nonumber \\
&&+\,\varphi_2(gA^\mu\partial_\mu)\varphi_1
+\varphi_1(-gA^\mu\partial_\mu)\varphi_2  \\
&&+\,a^\mu(2g^2A_\mu\phi)\varphi_1
+\,a^\mu(2g\partial_\mu\phi)\varphi_2 \nonumber \\
&&+\eta_1\frac{1}{2}\left(-\Box- g^2\phi^2\right)\eta_1
+\eta_2\frac{1}{2}\left(-\Box-g^2\phi^2\right)\eta_2
\nonumber  
\kma\end{eqnarray}
Here $\varphi_1$ and $\varphi_2$ denote the real and imaginary part of
the Higgs field fluctuations, $\eta_i$ are the Faddeev-Popov ghosts, and
we have chosen the 't Hooft-Feynman background gauge.
In compact notation this may be written as
 \be
\call^{II} =  \frac{1}{2}  \psi^*_i {\cal M}_{ij}  \psi_j \pkt
\end{equation}
The fields $\psi_i$ denote the ensemble of gauge, Higgs and
Faddeev-Popov fields and  the
fluctuation operator $\calm_{ij}$ is defined by this and the
previous equation.
In terms of the fluctuation operators
${\cal M}$ on the vortex and ${\cal M}^0$ for the
vacuum background fields, the effective action
is defined as
\begin{equation}
S_{eff} = \frac{i}{2} \ln \left\{ \frac{\det {\cal M}+i\epsilon}
{\det {\cal M}^0+i\epsilon} \right\} \; .
\end{equation}
As the background field is time-independent and also independent of
$z$ the fluctuation operators take the form
\be\label{flucsep}
\calm_{ij} = (\partial_0^2-\partial_3^2)\delta_{ij} +\calm_{\perp ij}
\kma \ee
where $\calm_\perp$ is a positive-definite operator describing the transversal
fluctuations.
It is identical for the longitudinal and timelike gauge fields and for
the Faddeev-Popov ghosts, so these contributions to the effective
action cancel. The remaining degrees of freedom form a
coupled system of four fields $\psi_i$: the real and imaginary part of 
the Higgs field fluctuations $\varphi_1,\varphi_2$ and the transverse 
gauge field fluctuations $a_1,a_2$.
 
As is well known the  logarithm of the determinant can be written as the
trace  of the logarithm. One can do the trace over $k_0$, the
momentum associated with the time variable, by integrating over
$T\int dk_0/2\pi$, where $T$ is the lapse of time.
One then obtains
\begin{equation}
S_{\rm eff} = -T \frac{1}{2} \sum \left[E_\alpha-E_\alpha^{(0)}\right]\kma
\end{equation}
where $E_\alpha$ are square roots of  the eigenvalues of the 
positive definite operator
\be
-\partial_3^2+\calm_\perp\kma
 \ee
and likewise $E_\alpha^{(0)}$ are those of the analogous operator
in the vacuum
\be
-\partial_3^2+\calm_\perp^0=-\partial_3^2-\vec \nabla_\perp^2 + {\bf m^2}
\pkt \ee
Here ${\bf m}^2=\diag(m_1^2,\dots,m_n^2)$ is the diagonal mass squared 
operator for the various fluctuations.

  So the effective action is  equal to the sum of differences between 
the zero point energies of the quantum
fluctuations around the vortex and the ones of the
quantum fluctuations in the vacuum, multiplied by $-T$.
Further, we can  do the trace over the variable $k_3$
by integrating over $L\int dk_3/2\pi$. We then obtain
\be
S_{\rm eff}= -TL\sum_\alpha\int\frac{dk_3}{2\pi}
\frac{1}{2}
\left[\sqrt{k_3^2+\mu_\alpha^2}-\sqrt{k_3^2+\mu_\alpha^{(0)~ 2}}\right]
\kma \ee
where $\mu_\alpha^2$ are the eigenvalues of the operator $\calm^\perp$ and
$\mu_\alpha^{(0)~ 2}$ those of $-\vec \nabla_\perp^2+{\bf m}^2$. 
In the same way the classical action becomes
\be
S_{\rm cl}=-TL\sigma_{\rm cl}
\ee
where $\sigma_{\rm cl}$ is the classical string tension.
The fluctuation part of the string tension is 
given by
\be \label{fluctuationstringtension}
\sigma_{\rm fl}=\sum_\alpha\int\frac{dk_3}{2\pi}
\frac{1}{2}\left[\sqrt{k_3^2+\mu_\alpha^2}-\sqrt{k_3^2+\mu_\alpha^{(0)~ 2}}\right]
\pkt \ee
Of course all expressions are formal, the integrals do not exist before
a suitable regularization. 

A way of computing $\sigma_{\rm fl}$ has been formulated in Ref.
\cite{Baacke:2008sq}: one computes the matrix valued 
Euclidian Green' s function of the
fluctuation operator defined by
\be
(p^2{\bf 1} +\calm_\perp) \calg(\bfx_\perp,\bfx'_\perp,p) =
{\bf 1}\delta^2(\bfx_\perp-\bfx'_\perp)
\ee
and similarly for $\calm_0$.
Then with
\be \label{fpdef}
F(p)=\int d^2 x_\perp \tr \left[\calg(\bfx_\perp,\bfx_\perp,p)-
\calg_0(\bfx_\perp,\bfx_\perp,p)\right]
\ee
the fluctuation correction to the string tension is given by
\be\label{sigmanum}
\sigma_{\rm fl}=-\int_0^\infty \frac{p^3~dp}{4\pi} F(p)
\ee
The function $F(p)$ has been computed in Ref. \cite{Baacke:2008sq}.
For small $p$ it behaves as $2/p^2$, as expected for 
two translation modes which yield  poles at $p=0$.
This is displayed in Fig. \ref{fig:zeromode} for
$\xi= m_H/m_W = 1$. For large momenta $F(p)$ again behaves
as $p^{-2}$, but with a different coefficient.
The contribution of the translation modes, 
as well as the entire fluctuation correction
are quadratically divergent. The renormalization has been
discussed previously \cite{Baacke:2008sq}.
We come back to the fluctuation correction below,
but at first we discuss  explicitly the translation mode
which describes collective oscillations of the vortex.

%%%%%%%%%%%%%%%%%%%%%%%%%%%%%%%%%%%%%%%%%%%%%%%%%%%%%%%%%%%%%%%%%%%%%%%%%

\section{Collective string oscillations}
\label{sec:collectivestringoscillations}
\setcounter{equation}{0}

For quantum kinks $\phi(x)$ the translation mode is proportional to the
derivative of the classical solution, $\psi_0= N d \phi_{\rm cl}(x)/dx$.
It is an eigenstate of the fluctuation operator with eigenvalue
zero, and it leads to a pole in the Green's function of the fluctuation
operator at energy $\omega=0$. For the quantum kink the translation mode is 
related to the collective motion of the entire kink, and
its contribution to the quantum corrections is the kinetic energy.
Here we are considering local transverse displacements of the vortex.
Each slice between $z$ and $z+\Delta z$ is moving separately and the
resulting motions of the vortex can be described in terms of
waves propagating along the string. In the Green's function
of the complete fluctuation operator $\calm$ the pole appears as a cut,
starting at $\omega=0$.

Of course we again expect the zero modes to be related to the derivatives
of the classical solution $\nabla_i \phi_{\rm cl}$, but the fluctuation 
operator is matrix valued and therefore we have to determine all
four components of the eigenvector. This is discussed in Appendix
\ref{app:translationmode}, in a cylindrical basis of modes for which
the fluctuation operator was derived in Refs. 
\cite{Baacke:1994bk,Baacke:2008sq}. 
The wave functions of the zero modes arising from local translation invariance
are derived in Appendix \ref{app:translationmode}.
Combining the modes with azimutal quantum numbers $m=\pm 1$
proportional to $\exp{\pm i\varphi}$ one finds that an infinitesimal
shift in the $x$ direction generates a four component wave function
\bea
\varphi^x_1(r,\varphi)&=&v f'(r) \cos \varphi \kma
\\
\varphi^x_2(r,\varphi)&=&-v \frac{A(r)+1}{r}f(r)\sin \varphi\kma
\\
{\bf a}^x(r,\varphi)&=& \left(\begin{array}{c}0\\-1\end{array}\right)
\frac{A'(r)}{gr}\kma
\eea
and a shift in the $y$ direction leads to
\bea
\varphi^y_1(r,\varphi)&=&v f'(r) \sin \varphi\kma
\\
\varphi^y_2(r,\varphi)&=&v  \frac{A(r)+1}{r}f(r) \cos \varphi\kma
\\
{\bf a}^y(r,\varphi)&=& \left(\begin{array}{c}1\\0\end{array}\right)
\frac{A'(r)}{gr}\pkt
\eea
The norm of these wave functions is given by
\bea \nonumber
||\psi_t^x||^2&=&\int r~dr~\int d\varphi
\left\{(a^x_1)^2+(a^x_2)^2+(\varphi^x_1)^2+(\varphi^x_2)^2\right\}
\\\nonumber&=&2 \pi v^2\int r~dr~\left\{\frac{1}{m_W^2}\frac{A'{}^2(r)}{r^2}+
\frac{1}{2}\frac{(A(r)+1)^2}{r^2}f^2(r)+\frac{1}{2}f'{}^2(r)\right\}
\kma\eea
and analogously for the $y$ mode. Using the virial theorem proven in
Appendix \ref{app:virialtheorem} it takes the value
\be
||\psi_t^x||^2 = \sigma_{\rm cl}
\kma
\ee
where $\sigma_{\rm cl}$ is the 
classical string tension. 

In the mode expansion of the quantum fields these modes appear
in the form
\bea \label{fieldexpansion1}
\phi_i(r,\varphi,z,t)=\phi^{\rm cl}(r,\varphi)\delta_{i1}+ 
\bfX(z,t)\varphi^x_i(r,\varphi)
+ \bfY(z,t)\varphi^y_i(r,\varphi) + \dots &&
\\\label{fieldexpansion2}
A_i(r,\varphi,z,t)=A_i^{\rm cl}(r, \varphi)+
\bfX(z,t)a^x_i(r,\varphi)
+\bfY(z,t)a^y_i(r,\varphi) + \dots&&
\kma\eea
where the dots indicate the contributions of all other
eigenfunctions of the fluctuation operator. While these have,
in general, complex wave functions we have written the contributions
of the translation modes, which are real, in a suggestive form
using operators $\bfX(z,t)$ and $\bfY(z,t)$. 
The canonical momenta of the field operators are given by
\bea
\Pi^\Phi_i(r,\varphi,z,t)=
\dot\Phi_i(r,\varphi,z,t = \dot\bfX(z,t)\phi^x_i(r,\varphi) + 
\dot\bfY(z,t)\phi^y_i(r,\varphi) + \dots && 
\\
\Pi^A_i(r,\varphi,z,t)=\dot A_i(r,\varphi,z,t)=\dot\bfX(z,t)a^x_i(r,\varphi)
+\dot\bfY(z,t)a^y_i(r,\varphi) + \dots &&
\kma\eea
 The relation of the operator $\bfX$ to the
usual creation and annihilation operators is given by
\bea\label{xcrelation}
\bfX(z,t)&=&\frac{1}{\sqrt{\sigma_{\rm cl}}}
\int \frac{dk}{2\pi 2|k|}\left(c_x(k)e^{i(kz-|k|t)}+
c_x^\dagger(k)e^{-i(kx-|k|t)}\right) \kma
\\\label{pcrelation}
\bfP_x(z,t)&=&\sqrt{\sigma_{\rm cl}} \int \frac{dk}{2\pi 2i}\left(c_x(k)e^{i(kz-|k|t)}-
c_x^\dagger(k)e^{-i(kx-|k|t)}\right)
\kma\eea
and analogously for $\bfY$. Here we have used the fact that for the zero 
modes we have $\omega=|k|$. 
The operators $c_\alpha(k)$ satify the commutation relations
\be
\left[c_\alpha(k),c_\beta^\dagger(k')\right]=
2\pi 2|k|\delta_{\alpha\beta}\delta(k-k')
\kma\ee
and the commutation relation between $\bfX$ and $\bfP_x$ is
\be
\left[\bfX(z,t),\bfP_x(z',t)\right]= i \delta(z-z')
\pkt\ee
 The normalization factors $\sqrt\sigma_{\rm cl}$
in Eqs. \eqn{xcrelation} and \eqn{pcrelation} are determined by 
the requirement that in the field 
expansion  the operators $c_x(k), c_x^\dagger(k)$ have to appear multiplied
by wave functions normalized to unity. 
This is necessary for obtaining the 
canonical equal time commutation relations for the fields,
\bea
\left[\Phi_i(x,y,z,t),\dot \Phi_j(x',y',z',t)\right]=i\delta_{ij}
\delta^3(\bfx-\bfx')
\\
\left[A_i(x,y,z,t),\dot A_j(x',y',z',t)\right]=i\delta_{ij}
\delta^3(\bfx-\bfx')
\eea
via the completeness relation of the wave functions. Of course, this 
completeness relation requires of the inclusion of all
eigenfunctions of the fluctuation operator, which 
above are indicated by dots.

The second order 
Hamilton operator corresponding to the Lagrangian
\eqn{lag2} can be written in the form
\be
\bfH^{II}=\int d^2 x_\perp \int dz
\frac{1}{2}\left\{\sum_i \dot \psi_i^2+\sum_i\left[\frac {d\psi_i}{dz}\right]^2
+\sum_{ij}\psi_i \calm_{\perp ij}\psi_j\right\}
\pkt\ee
Here we are interested only in the contribution of the zero modes
of $\calm_\perp$.
If we insert the fluctuation fields of the
field expansion Eqs. \eqn{fieldexpansion1} and \eqn{fieldexpansion2}
the operator $\calm_{\perp ij}$ does not contribute. Using further the
norm of the translation modes in order to do the integration
over $d^2 x_\perp$ we obtain
\be
\bfH^{II}_{\rm transl.}=
\sigma_{\rm cl}\int dz \frac{1}{2}\left\{
\dot \bfX^2+ \bfX'{}^2+\dot \bfY^2+ \bfY'{}^2\right\}
\pkt \ee
Including the classical string tension we find
\be \label{Heffnonrel}
\bfH=\sigma_{\rm cl} \int dz \left[1+\frac{1}{2}\left(
\dot \bfX^2+ \bfX'{}^2+\dot \bfY^2+ \bfY'{}^2\right)\right\}+\dots
\kma\ee
where the dots indicate the contributions of higher modes.
This looks analogous to the result for the quantization of kinks
\be
H= M+\frac{1}{2M}\bfP^2+\dots= M(1+\frac{1}{2}\dot\bfX^2)+\dots
\pkt\ee
There we know that the complete result, which only appears if higher loops
are included, must be Lorentz covariant:
\be
H= \frac{M}{\sqrt{1-\dot\bfX^2}}+\dots
\kma\ee
corresponding to an action
\be
S=-M\int dt \sqrt{1-\dot \bfX^2}
\pkt
\ee
For the case of the Nielsen-Olesen vortex the action
\be \label{Seffrel}
S=-\sigma_{\rm cl}\int dt\int dz\sqrt{1-\dot \bfX^2-\dot \bfY^2+
\bfX'{}^2+\bfY'{}^2}
\pkt\ee
implies the string Hamiltonian
\be \label{Heffrel}
H=\sigma_{\rm cl} \int dz \frac{1+\bfX'{}^2+\bfY'{}^2}
{\sqrt{1-\dot \bfX^2-\dot \bfY^2+
\bfX'{}^2+\bfY'{}^2}}
\ee
which in the nonrelativistic limit leads to Eq. \eqn{Heffnonrel}.
In this limit these results are in agreement with Ref. \cite{Nielsen:1973cs}

%%%%%%%%%%%%%%%%%%%%%%%%%%%%%%%%%%%%%%%%%%%%%%%%%%%%%%%%%%%%%%%%%%%%%%%%%

\section{Energy of collective fluctuations and renormalization}
\label{sec:fluctuationenergy}
\setcounter{equation}{0}
The Hamiltonian for collective oscillations of the vortex
primarily describes excitations of the string, here:
transversal waves that propagate along the $z$ axis.
The zero point energies associated with these degrees of freedom
can be absorbed, in the string picture,
 into the redefinition of the string tension.
In quantum field theory they are absorbed, as all other divergences,
by counter terms local in the fields, the string tension does not appear
in the basic Lagrangian and there is no related counter term either.
The difference between the two approaches appears in a similar
way in the case of the Casimir effect \cite{Baacke:1985fh,Graham:2002fi}.
 We would like to discuss this in some detail. 

\begin{figure}
\begin{center}
\includegraphics[scale=0.5]{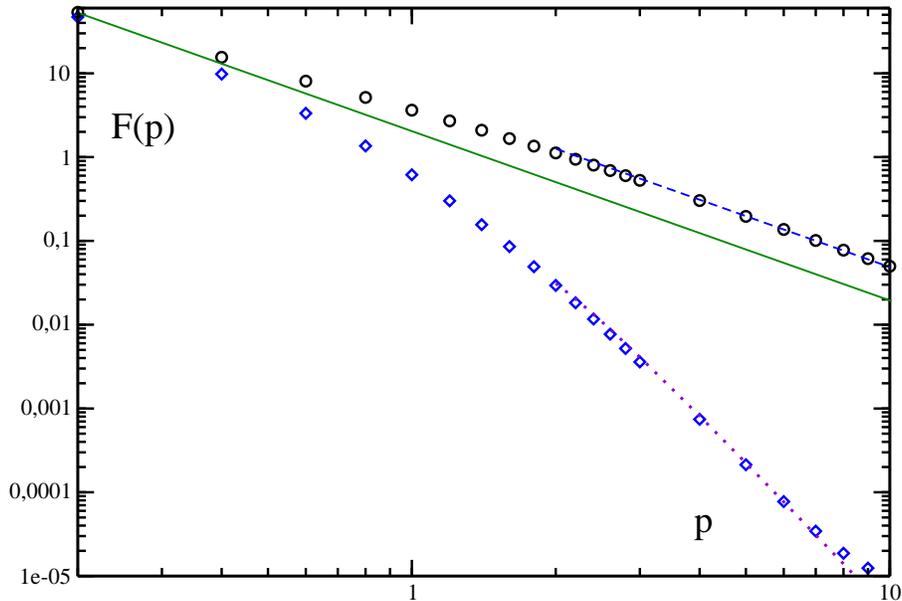}
\end{center}
\vspace{4mm}
\caption{\label{fig:zeromode}
The integrand function $F(p)$ defined in Eq.
\eqn{fpdef}: circles: the unsubtracted function;
dashed line: asymptotic behaviour $ a/p^2$;
solid line: zero mode contribution $2/p^2$;
diamonds: the subtracted function; dotted line: asymptotic
behaviour $\propto 1/p^6$ of the subtracted function. }
\end{figure}

The trace of the  Euclidian  Green' s function 
$F(p)$ for the gauge-Higgs sector is displayed in Fig.
\ref{fig:zeromode}. We have mentioned
already that at low momenta it behaves as $2/p^2$ which is the
reflection of the two zero modes. At high momenta it behaves
as $a/p^2+b/p^4+ O\left(p^{-6}\right)$ where the coefficients
$a$ and $b$ are determined  by the lowest orders 
of perturbation theory. A contribution to the Green' s function 
which at high momenta is proportional to $1/p^2$ is converted, 
via Eq. \eqn{sigmanum} into a quadratic divergence for the 
one-loop string tension. 
If we consider the complete Green' s function then this quadratic
divergence, as well as the subleading logarithmic one,
can be handled \cite{Baacke:2008sq} by subtracting 
the leading orders perturbation theory
analytically and by regularizing and renormalizing them in the
usual way. The subtracted function, whose integral is finite,
behaves as $p^{-6}$; it is plotted in Fig. \ref{fig:zeromode}.
The zero mode poles are part of the asymptotic
Green' s function; if they were removed, the asymptotic behaviour
of the subtracted Green's function would be $-2/p^2$ and its
integral would again be divergent. 
We neither have a prescription to handle this divergence
nor the one of the separated zero mode pole. 
So within usual renormalized perturbation
theory there is no obvious way to quantify the contribution of the
collective string oscillations to the total one-loop fluctuation
energy: the renormalization of the contribution of collective
fluctuations to the string tension is embedded into
the renormalization of the entire one-loop contribution within the
framework of renormalized quantum field theory. It is not necessary to
invoke a mathematical definition of divergent sums like the
zeta function regularization.

There is further a conceptual difference between the zero point energy
of the collective fluctuations in a string picture
 and their contribution to the one-loop corrections
in quantum field theory:
In a pure string picture the presence of fluctuations
trivially requires the presence of the string. Once included
their zero point energies are added to the string tension. 
So, if their contribution were be finite it would positive.
In quantum field theory the fluctuations of the field are present
even in the absence of the vortex. The vortex generates an
attractive potential. The presence of the
zero modes implies that levels of a
continuum which starts at energies larger than $m_W$ or $m_H$
are pulled down such that at least in
one channel the continuum starts at energy zero. So we expect
a negative contribution to the string tension. Indeed the
unsubtracted  function $F(p)$ is positive and if the
integral of Eq. \eqn{sigmanum} were finite, $\sigma_{\rm fl}$
would be negative. This simple
feature gets obscured in the process of regularization and 
renormalization.

%%%%%%%%%%%%%%%%%%%%%%%%%%%%%%%%%%%%%%%%%%%%%%%%%%%%%%%%%%%%%%%%%%%%%%%%%

\section{Conclusions}
\setcounter{equation}{0}
\label{sec:conclusions}
We have considered here a particular aspect of the one-loop
corrections to the string tension of the Nielsen-Olesen vortex,
the r\^ole of the translation modes. We have identified their wave functions
and derived their contribution to the string tension. This contribution
describes the energy of transversal waves propagating along the
direction of the string. These can be considered
as collective fluctuations of the classical vortex, in the same way
as the translation modes of quantum kinks describe the collective
motion of the kink. This relation is made precise, in both cases,
by virial theorems. The effective action for the
fluctuations is found to be the nonrelativistic limit
of the Nambu-Goto action.

For the handling of the divergent one-loop corrections
we have discussed conceptual differences between standard
string theory and the vortex of the Abelian Higgs model.
In the latter case the divergences associated 
with the the zero point energies of collective
fluctuations are treated along with those of other
fluctuations within the standard framework
of renormalized quantum field theory.

The approach described here only pertains to a
straight line vortex of infinite length. It can be expected
to hold for more general string configurations as long as the
curvature radii and a possible finite length are large
compared with the transverse extension of the string. The 
conceptual differences in renormalization between 
the idealized string model
and the vortex of a quantum field theory are of course
of an entirely general nature. Indeed they are more general than
the special model considered here. E.g., if we elevate a kink
of an $1+1$ dimensional quantum field theory to a domain wall in $3+1$ 
dimensions, its translation mode reappears in the form of collective
surface oscillations and the renormalization of this degree of freedom
is again embedded into the renormalization of the energy
of all quantum fluctuations.

%%%%%%%%%%%%%%%%%%%%%%%%%%%%%%%%%%%%%%%%%%%%%%%%%%%%%%%%%%%%%%%%%%%%%%%%%

\section*{Acknowledgements}
The author has pleasure in thanking Nina Kevlishvili for
useful discussions and comments.

%%%%%%%%%%%%%%%%%%%%%%%%%%%%%%%%%%%%%%%%%%%%%%%%%%%%%%%%%%%%%%%%%%%%%%%%%

\newpage
\noindent
{\LARGE \bf Appendix}
\setcounter{equation}{0}

\begin{appendix}

%%%%%%%%%%%%%%%%%%%%%%%%%%%%%%%%%%%%%%%%%%%%%%%%%%%%%%%%%%%%%%%%%%%%%%%%%

\section{The translation mode}
\setcounter{equation}{0}
\label{app:translationmode}
The fluctuation operator  for the coupled system
of transverse gauge and Higgs fields
was derived in Ref. \cite{Baacke:2008sq}
in a basis of partial waves with ``magnetic'' quantum numbers
$m$, proportional to $\exp(im\varphi)$. We refer to
this reference for details. Essentially, the amplitudes
$F_4^m$ corresponds to the real part of the Higgs field
$\varphi_1$, $F_3^m$ to the imaginary part of the
Higgs field $\varphi_2$, and the amplitudes $F_1^m$ and $F_2^m$ to
combinations of the transverse gauge fields $a_1,a_2$. The
basis was chosen in such a way that the fluctuation operator
becomes symmetric, and the amplitudes are real relative to each other.
The amplitudes $F_i$ for $m= 1$ satisfy the following 
coupled system of linear
differential equations:
\bea
\left\{-\frac{1}{r}\frac{d}{dr}r\frac{d}{dr}+m_W^2 f^2\right\}
F^1_1+\sqrt{2}m_W f' F^1_3+\sqrt{2}m_W\frac{A+1}{r}f F^1_4=0&&
\\
\left\{-\frac{1}{r}\frac{d}{dr}r\frac{d}{dr}+\frac{4}{r^2}+m_W^2 f^2\right\}
F^1_2+\sqrt{2}m_W f' F^1_3-\sqrt{2}m_W\frac{A+1}{r}f F^1_4=0&& 
\\\nonumber
\left\{-\frac{1}{r}\frac{d}{dr}r\frac{d}{dr}+\frac{1}{r^2}+
\frac{(A+1)^2}{r^2}+m_W^2 f^2+\frac{m_H^2}{2}\left(f^2-1\right)\right\}
F^1_3&&
\\+\sqrt{2}m_W f'(F^1_1+F_2)-2\frac{A+1}{r} F^1_4=0&& 
\\
\nonumber
\left\{-\frac{1}{r}\frac{d}{dr}r\frac{d}{dr}+\frac{1}{r^2}+
\frac{(A+1)^2}{r^2}+\frac{m_H^2}{2}\left(3f^2-1\right)\right\}
F^1_4&&
\\+\sqrt{2}m_W\frac{A+1}{r}(F^1_1-F^1_2)-2\frac{A+1}{r} F^1_3=0&& 
\pkt\eea
The last equation corresponds to the real part of the fluctuations
of the field $\phi$, and the translation mode is obtained
as $\nabla \phi_{cl}=\hat x v f'$. We therefore start with the ansatz
\be
F^1_4= c f'
\kma\ee
where the coefficient $c$ is a prefactor that will be fixed
later. Applying the derivative $d/dr$ to the equation of motion
for $F_4^1$ we find
\bea\nonumber
\frac{d}{dr}\left\{-\frac{1}{r}\frac{d}{dr}r\frac{d}{dr}
+\frac{\left[A(r)+1\right]^2}{r^2}+\frac{m^2_H}{2}\left[f^2(r)-1\right]
\right\}f(r)&=&
\\\nonumber
\left\{-\frac{1}{r}\frac{d}{dr}r\frac{d}{dr}
+\frac{\left[A(r)+1\right]^2}{r^2}+\frac{m^2_H}{2}\left[3f^2(r)-1\right]
\right\}f'&&
\\
-2\frac{(A+1)^2}{r^3}f+2\frac{A+1}{r^2}fA'&=&0
\pkt\eea
This is the  equation of motion for $F_1$ if the choose
\bea
F_3^1=c\frac{A+1}{r}f
\\
F_1^1-F_2^1=c \frac{\sqrt{2}}{m_W}\frac{A'}{r}
\pkt
\eea
The assignment has to be checked for consistency with the
remaining equations of motion. Multiplying the equation of motion of 
$f(r)$ with $(A+1)/r$ and commuting this factor with 
the derivatives we find
\bea\nonumber
\frac{A+1}{r}\left\{-\frac{1}{r}\frac{d}{dr}r\frac{d}{dr}
+\frac{\left[A(r)+1\right]^2}{r^2}+\frac{m^2_H}{2}\left[f^2(r)-1\right]
\right\}f(r)&=&
\\\nonumber
\left\{-\frac{1}{r}\frac{d}{dr}r\frac{d}{dr}
+\frac{1}{r^2}+\frac{\left[A(r)+1\right]^2}{r^2}+\frac{m^2_H}{2}\left[f^2(r)-1\right] + m_W^2 f^2
\right\}\frac{A+1}{r}f(r)&&
\\
-2\frac{(A+1)^2}{r^2}f'+2f'\frac{A'}{r}=0&&
\pkt\eea
In the intermediate steps we have used the equation of motion
for $A(r)$ in order to replace a second derivative $A"$.
The result is consistent with the previous assignement if we
choose
\be
F_1^1+F_2^2=c \frac{\sqrt{2}}{m_W}\frac{A'}{r}
\pkt\ee
So we find
\bea
F_2^1&=&0
\\
F_1^1&=&c \frac {\sqrt{2}}{m_W}\frac{A'}{r}
\pkt\eea 
This has to be verified by deriving the equation of motion for
$A'/r$. We obtain it by applying $1/r~d/dr$ to the 
classical equation of motion for $A(r)$:
\bea\nonumber
\frac{1}{r}\frac{d}{dr}\left\{
-r \frac{d}{dr}\frac{1}{r}\frac{d}{dr}
+m^2_W f^2\right\}
\left[A+1\right]&=&
\\
\left\{-\frac{1}{r}\frac{d}{dr}r \frac{d}{dr}+m_W^2 f^2\right\}
\frac{A'}{r}+2m_W^2\frac{A+1}{r} ff'=0&&
\pkt
\eea
This is consistent with the previous assigments and the equation
of motion for $F_1^1$.

So we have derived that the wave function of the translation
mode with $m=1$ is given by
\be
\left(\begin{array}{c}F_1^1\\F_2^1\\F_3^1\\F_4^1\end{array}
\right)=c\left(\begin{array}{c}
\sqrt{2}/m_W~A'/r\\
0\\
(A+1)f/r\\
f'\end{array}\right)
\ee
We have started this derivation by considering the gradient
applied to the scalar field $\phi=v f(r)$, but this has fixed
only the component $F_4$. The component $F_3$ is easily seen
as arising from replacing the derivatives $\nabla_i$ by the
covariant derivatives $\nabla_i-ig A_i$. The wave function for the 
vector potential is not given by an infinitesimal shift of the
classical potential, but it correctly describes the shift in
the classical magnetic field.

There is of course a second translation mode, with $m=-1$. Its wave function
is given by
\be
\left(\begin{array}{c}F_1^{-1}\\F_2^{-1}\\F_3^{-1}\\F_4^{-1}\end{array}
\right)=c'\left(\begin{array}{c}
0\\
\sqrt{2}/m_W~A'/r\\
-(A+1)f/r\\
f'\end{array}\right)
\ee
From these wave functions in the azimutal basis and in terms
of the amplitudes $F_i^{\pm 1}$ we go back to the physical basis
$\varphi_1,\varphi_2,a_1,a_2$.
These are related to the functions $F_i^1$
via \footnote{The prefactors appearing in the definitions
of $F_1$ and $F_2$, Eq. (4.2) of Ref. \cite{Baacke:2008sq}
should be $1/\sqrt 2$, not $1/2$. This misprint also appears
in Refs. \cite{Baacke:1994bk} and \cite{Baacke:2008zx}}
\be
\left(\begin{array}{c}a_1\\a_2\\\varphi_1\\\varphi_2\end{array}
\right)=\frac{1}{\sqrt{2\pi}}\left(\begin{array}{c}
(F_1^1+F_1^{1*})/\sqrt{2}\\
i(F_1^1-F_1^{1*}/\sqrt{2}\\
-i[F_4^1\exp(i\varphi)-F_4^{1*}\exp(-i\varphi)]\\
F_3^1\exp(i\varphi)+F_3^{1*}\exp(-i\varphi)
\end{array}\right)
\ee
Here we have used reality constraints for the fields in order
to eliminate the functions $F_i^{-1}$.
The wave functions corresponding to infinitesimal translations
into the $x$ and $y$ directions are obtained by
choosing the free coefficient $c$ is such a way that the real part of the
Higgs field fluctuation $\varphi_1$ is given by $d \phi_{\rm cl}/dx$
and $d \phi_{\rm cl}/dy$, respectively. Explicitly, 
$c_x=iv\sqrt{\pi/2}$ and $c_y=v \sqrt{\pi/2}$.
The complete wave functions  are 
given in section \ref{sec:collectivestringoscillations}.

%%%%%%%%%%%%%%%%%%%%%%%%%%%%%%%%%%%%%%%%%%%%%%%%%%%%%%%%%%%%%%%%%%%%%%%%%

\section{The virial theorem}
\label{app:virialtheorem}
\setcounter{equation}{0}
As we have discussed in section \ref{sec:collectivestringoscillations}
the normalization of the translation mode derived from the
deformation of the classical solution
is given by
\be
|\psi_t|^2=\pi v^2\int r~dr~\left\{\frac{2}{m_W^2}\frac{A'{}^2(r)}{r^2}+
\frac{(A(r)+1)^2}{r^2}f^2(r)+f'{}^2(r)\right\}
\ee
The classical string tension is given by Eq. \eqn{eq:classtens},
i.e., 
\bea\nonumber
\sigma_{\rm cl}&=&\pi v^2\int r~dr~\left\{\frac{1}{m_W^2}\frac{A'{}^2(r)}{r^2}+
\frac{(A(r)+1)^2}{r^2}f^2(r)+f'{}^2(r)\right.
\\
&&\left.\hspace{25mm} +\frac{m_H^2}{4}\left[f^2(r)-1\right]^2\right\}
\eea
In analogy with the virial theorem for quantum kinks, where the normalization
of the translation mode is equal to the classical mass,
we expect a virial theorem $|\psi_t|^2=\sigma_{\rm cl}$ which
reduces to the identity
\be
\int r~dr~\frac{1}{m_W^2}\frac{A'{}^2(r)}{r^2}
=\int r~dr~\frac{m_H^2}{4}\left(f^2(r)-1\right)^2
\pkt\ee
It can readily be  verified numerically. In order
to derive the relation analytically we use the following weighted integrals 
over the classical equations of motion
\bea
&&I_1=\int r~dr~f\left\{
-\frac{1}{r}\frac{d}{dr}r\frac{d}{dr}
+\frac{\left[A+1\right]^2}{r^2}+
\frac{m^2_H}{2}\left(f^2-1\right)
\right\}f=0
\\
&&I_2=\int r~dr~f r\frac{d}{dr}\left\{
-\frac{1}{r}\frac{d}{dr}r\frac{d}{dr}
+\frac{\left[A+1\right]^2}{r^2}+
\frac{m^2_H}{2}\left(f^2-1\right)
\right\}f=0
\\
&&I_3=\int dr~(A+1)\frac{d}{dr}
\left\{-r\frac{d}{dr}\frac{1}{r}\frac{d}{dr}+m^2_W f^2\right\}
\left(A+1\right)=0 \hspace{25mm}
\eea
Integrating by parts these take the form
\bea 
I_1 &=&\int r~dr~\left\{\left(\frac{df}{dr}\right)^2
+\frac{(A+1)^2}{r^2}f^2+\frac{m_H^2}{2}f^2(f^2-1)\right\}=0 
\\\nonumber
I_2 &=&\int r~dr~\left\{-2\left(\frac{df}{dr}\right)^2
-2\frac{(A+1)^2}{r^2}f^2-m_H^2f^2(f^2-1)\right.
\\
&&\left.\hspace{25mm}+\frac{m_H^2}{4}(f^2-1)^2-\frac{(A+1)^2}{r}ff'\right\}=0 
\\
I_3 &=& \int r~dr~\left\{-\frac{A'{}^2}{r^2}
+m_W^2\frac{(A+1)^2}{r}ff'\right\}
\eea
Combining the first and second integral we find
\be
I_2+2I_1=\int r~dr~\left\{\frac{m_H^2}{4}(f^2-1)^2-
\frac{(A+1)^2}{r}ff'\right\}=0
\ee
Adding the third integral we obtain the relation
\be
I_2+2I_1+\frac{1}{m_W^2}I_3=\int r~dr~\left\{
-\frac{1}{m_W^2}\frac{A'{}^2}{r^2}+
\frac{m_H^2}{4}\left(f^2-1\right)^2\right\}=0
\ee
which is the expected result.
\end{appendix}

\bibliography{novqc}
\bibliographystyle{h-physrev4}

\end{document}